# AN INTEGRATED GEOGRAPHIC INFORMATION SYSTEM AND MARKETING INFORMATION SYSTEM MODEL

A. Quist-Aphetsi Kester, Lecturer Faculty of Informatics, Ghana Technology University; B. Dr. Koumadi, Koudjo M, Lecturer Graduate School, Ghana Technology University; C. Prof. Nii.Narku Quaynor, Professor, Computer Science Department, University of Cape Coast.

## Abstract

Maintaining competitive advantage is significant in this present day of globalization, knowledge management and enormous economic activities. An organization's future developments are influenced by its managements' decisions. Businesses today are facing a lot of challenges in terms of competition and they have to be in the lead by strengthening their research and development strategies with the aid of cutting edge technologies. Hence marketing intelligence is now a key to the success of any business in today's rapidly changing business environment. With all the technologies available in marketing research, businesses still struggle with how to gather information and make decisions in a short time and real-time about their customers' needs and purchasing patterns in various geographical areas.

This paper is set out to contribute to the body of knowledge in the area of the application of Geographic Information Systems technology solutions to businesses by developing a model for integrating Geographic Information Systems into existing Marketing Information Systems for effective marketing research. This model will interconnect organizations at the highest levels, providing reassurance to enable broad scope of checks and balances as well as benefiting many business activities including operational, tactical and strategic decision making due to its analytical and solution driven functions.

## Introduction

Marketing is the activity, set of institutions, and processes for creating, communicating, delivering, and exchanging offerings that have value for customers, clients, partners, and society at large. [1]

In the past few years, enhancement of technologies has stimulated several innovations in marketing research. These innovations include online survey methods and focus groups, e-commerce customer tracking systems, data mining tools, and 3-D graphics software. Technology is changing the way marketing data are collected, analyzed, and used for supporting managerial decisions. Marketing research provides managers with information that can be used to identify and define marketing opportunities and problems; generate, refine, and evaluate marketing actions; monitor marketing performance; and improve our understanding of marketing as a process [2].

According to the global information technology report 2012, there are more than 500 million mobile phone subscribers in Africa today, and Apple is the world's largest company in market capitalization, producing iPhones, iPods, and iPads along with Mac computers. Despite the strides the sector has made since the technology bust in 2001, however, we believe we are only just beginning to feel the impact of digitization the mass adoption by consumers, businesses, and governments of smart and connected ICT. Hence businesses have taken advantage of digitization businesses processes. This puts them in the lead of serving their customers across all geographical areas effectively. [3]

Today, more than 70 percent of the world's citizens live in societies that have just begun their digitization journeys. As the individuals and enterprises in these societies continue to progress in developing their own digitization capabilities, they will only increase and accelerate these economic and social benefits [3]. Hence businesses have to take advantage of this and digitize their market research activities with cutting edge technologies in other to meet the demands of the future. Data in data warehouses should be structured in such a way that data can easily and intelligently be mined in order to meet real-time business needs geographically.

Data is the lifeline of every company in today's world; organizations gather and continue to continue to gather huge quantities of data. The design and architecture of a data warehouse must be flexible enough to grow and change with the business needs. A data warehouse contains a copy of transaction data specifically structured for querying and analysis. It enables organizations to manage information as a whole. A critical tool for information and wisdom transformation is Business Intelligence. [4]

Throughout the history of systems development, the primary emphasis had been given to the operational systems and the data they process. The fundamental requirements of the operational and analysis systems are different: the operational systems need performance, whereas the analysis systems need flexibility and broad scope [5].

The combination of data from multiple source applications such as sales, marketing, production and finance manifests the effectiveness of a data warehouse system. Data warehouse systems are most successful when their design aligns with the general business structure rather than specific requirements. [6]

Management's access to the right information is among the most efficient ways of decision making. GIS as part of a marketing information system observed how the analytical and data incorporation strengths of GIS can be used in internal reports, marketing intellect systems, marketing decision support analysis and marketing exploration, with the overall aim of further understanding customer behavior. [7][8]

Modern day GIS reformation by Environmental System Research Institute (ESRI) has enabled the use of combining digital geographic features and aerial pictures that allow effective decision making. [9]

In today's global economy, the lines between industries and countries blur. Research often crosses many boundaries. Driven by rapid advances in technology, legislative reform, and global competition, the communications world is changing at an ever-accelerating rate. [9]

This study is intended to design a model to integrate the technology geographic information system GIS with marketing information systems MkIS to enhance productivity and competitive advantage. This model will interconnect organizations at the highest levels, providing reassurance to enable broad scope of checks and balances protecting executives as well as benefit many business activities including operational, tactical and strategic decision making due to its analytical and solution driven functions.

The paper has the following structure: section II consist of related works, section III gives information on the methodology, section IV Integration of GIS and MKIS, section V gives information regarding the architecture used, section VI provided a framework underlying the structure supporting the service-abstraction level and section VII concluded the paper.

## Related Works

A GIS is a systematized collection of computer hardware, software, geographic data, and personnel intended to efficiently capture, store, update, manipulate, analyze, and display all forms of geographically referenced information [8]. The impact of GIS application is beneficial in many business activities including: operational, tactical and strategic decision making levels due to its analytical and solution drive functions. The field of retail enhances these features to determine points of sales, retail sites, and analyze future and competitive development.

Since 1969, ESRI has aided more than 100, 000 organizations worldwide using advance technology for location information management. GIS integrates computers, software, and data to leverage the fundamental principle of geography that location is important to business. It enables you to observe, understand, question, analyze, visualize and interpret data from databases and spreadsheets onto a map, highlighting the locations and demographic profile of your customers from your store. [10]

Understanding the population dynamics around a resource can be a valuable tool for managers. Demographic analysis can highlight trends in the size and density of communities adjacent to a protected area, as well as factors attracting or discouraging individuals from residing in these areas. By linking population data and use patterns, demographic analysis can help managers target the sources of practices that are detrimental to the health of marine ecosystems. This type of information can be incorporated into a geographic information system (GIS) to display demographic patterns spatially, linking ecological and human characteristics within a landscape. [11]

Market Intelligence (MKI) is gathered through internal analysis, competition analysis, and market analysis about the total environment forming a broad spectrum of assembled knowledge, which is then used for developing scenarios so that timely reporting of vital foreknowledge for future planning in the areas of strategic, tactical, and counter-intelligence decision-making can be applied operationally and strategically in respect to the whole organization's strategic interest for the whole market.[12][13]

Due to the dynamic and volatile competitive environments, companies are now expanding their geographical market coverage. This increases management's quick demand for more information as income rises and buyers become more selective in their choice of goods.[14] In times past, MkIS was used as a decision support system for marketing management.

Data warehouse enables organizations to manage information as a whole. A critical tool for information and wisdom transformation is Business Intelligence [5]. A data warehouse can either be normalized or denormalized, a relational database, multidimensional database, flat file, hierarchical database, object database, etc. Data warehouse data is often changed. It often based on a specific activity or entity [15]. Companies that have made large investments in GIS have achieved considerable cost savings [16]. Despite numerous benefits of GIS, many retailers have been slow to investigate the possibilities of GISs [8] [17]. All managers are reluctant to incur the costs associated with the implementation of information systems, unless convinced of the benefits. Costs of implementing GISs, both in terms of initial capital investment for hardware and software and also in terms of its influence on organizational structures and approaches to management decision making can be high. Considering the costs of the systems themselves, additional staff in technical and managerial roles to manage them is also required.

Most features depicted on a map have attribute information, descriptive data about each feature. The attributes of a shopping mall, might include the name, type, size, names of anchor stores, a list of tenants, and the number of parking spaces available. However, with paper maps you can only display a limited amount of attribute information using the map symbols. For example, the width and color of the symbol used to depict a road can discriminate between roads, highways, freeways, and interstate highways. In GIS representations locations can be polygons (Fig. 1A), such as city, county, or sales territory boundaries; point features (Fig. 1B), such as customer street addresses, building locations, or vending machine locations; or linear features (Fig. 1C), such as roads, rivers, or railroads.

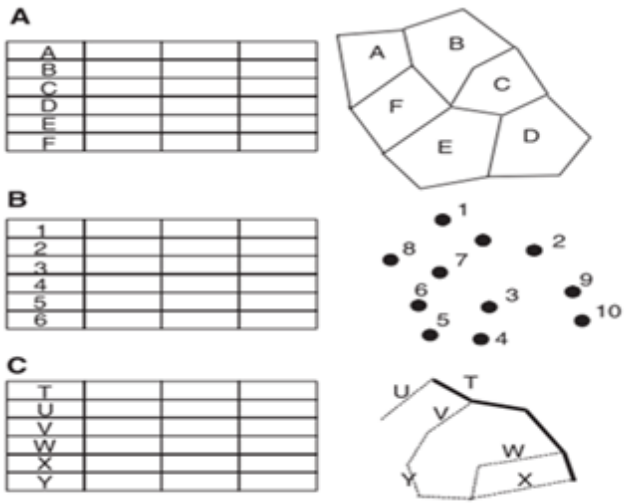

**Figure 1. GIS multiple themes**

The banking industry in recent years has been subjected to a variety of revolutions ranging from advancement in information technology, financial operations and intense competition among local banks, international banks and other financial institutions. Considering the challenges in the banking industry requires a technological or database upgrade that integrates geographical modeling into their existing system providing unique advantages to the banking community.

Data is the lifeline of every company in today's world; organizations gather and continue to continue to gather huge quantities of data. The design and architecture of a data warehouse must be flexible enough to grow and change with the business needs. A data warehouse contains a copy of transaction data specifically structured for querying and analysis.

Below are a few ways by which GIS will enhance banking business goals:
1. Expansion of Customer base
2. Improvement in Quality of Service
3. Consistent Business Growth
4. Increase in Profitability
5. Increase Customer Satisfaction

The aid of GIS in the banking industry can be appreciated in diverse areas especially in strategic planning and decision making.

## Methodology

The interest of GIS keeps rising world-wide. Managers and decision makers need to integrate the GIS technology into business systems to help them analyze data effectively. The application of GIS within the marketing research industry is an emerging even though GIS technology has been in existence in the 1990's. The many advantages of GIS include the possibility of integrating spatial and alphanumeric data making it widely applicable to a variety of fields [17] GIS also allows retailers, and virtually any business organization, to go beyond data integration and map generation to explore relationships within a wide range of data [8] [18]. It is therefore necessary for organizations to leverage the data and applications developed as part of this GIS project in their retail outlets to enable them compete, maintain and win market share [9] The combination of Service Oriented Architecture and Web services can be used to provide a rapid integration solution for MKIS applications that are residing on legacy systems and for already existing MKIS systems that will not be compatible with the integration with the GIS server. This will quickly and easily align Information Technology investments and corporate strategies by focusing on shared data and reusable services rather than proprietary integration products of MKIS systems or Information systems holding transactional data of organizations The GIS server and the MKIS system can further be integrated into a service bus [19].

This study aims to investigate this new phenomenon through the integration of GIS and MkIS to enable the deeper understanding of the dynamics of this industry and then propose a model for its effective integration. This study focuses on the design of architectures to illustrate the model of integration between MkIS and GIS within the marketing research industry.

## The GIS and MKIS Integration Model

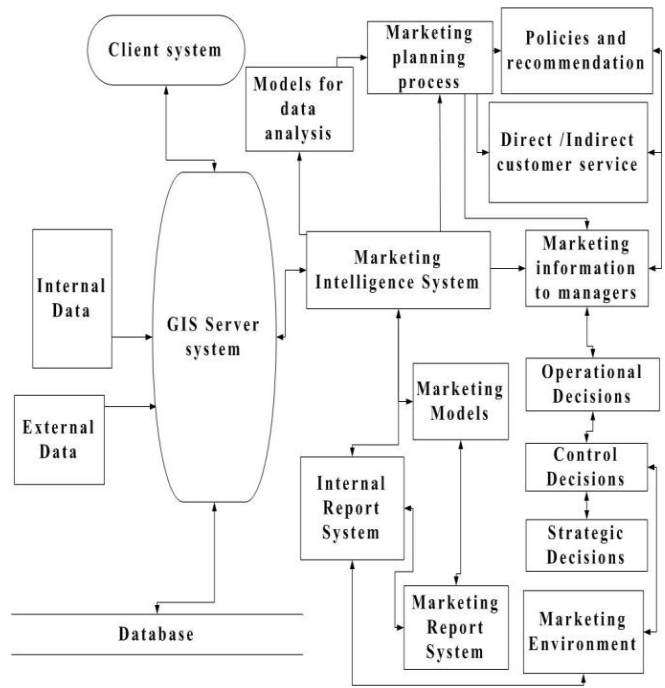

**Figure 2. GIS and MkIS integration Model**

The GIS server provides services such as the map service, and address locators, geodata service, and image service of locations of various transactions. The GIS server hosts resources and exposes them as services to client applications. The GIS Server System processes Client request to and from the Client System. The GIS Server Operator models structure of data definition for analysis. Client systems processes are integrated with the GIS server to track transaction locations of users and keep transaction data in the database. The GIS Server then stores and retrieves from the Database and forwards it to the Marketing Intelligence System for further processing whenever such data is needed. The GIS server is also connected to external and internal data sources of other distributed systems to enhance analyses of relationships between datasets within GIS enable systems and non GIS enabled systems by mining data and extracting information from it for effective decision making.

The Marketing Intelligence system retrieves data from the GIS server and forwards it to other units within the architecture for usage. The Models for data analysis creates models based on predefined system and this further forwarded to marketing planning process unit. Effective marketing strategies are devised from the available data and models for the formulation of policies and management action. This data can also be used to do future forecasting, strategic decision making.

phones, Management Server, and Institutions) via the Internet. It also enables the various components access to demographic, spatial and lifestyle information to enhance decision making. E-commerce Server allows businesses to connect to their customers via the GIS server by providing demographic information of the customers to enhance productivity and competitive advantage. It also allows customers to be able to acquire geographic data on their suppliers.

Smart phones and mobile devices enable users to connect directly to the GIS server via the internet on the go. This provides access to limitless geographic, consumer and business information to enhance decision making.

Management Server enables management to connect via the internet to the GIS server and query various kinds of information such as: market trends, geospatial data and business information to promote competitive advantage and facilitate decision making.

Financial Institutions and others can connect to the GIS server to access monitor business trends, geographic and demographic customer data to enhance productivity and decision making. This aids in identifying the appropriate location for new branch sites and business opportunities to gain competitive advantage.

Local government city planners access geospatial data from the GIS server to aid proper planning of the City. These data will promote: waste management, road networks and identify disaster prone areas to enhance decision making.

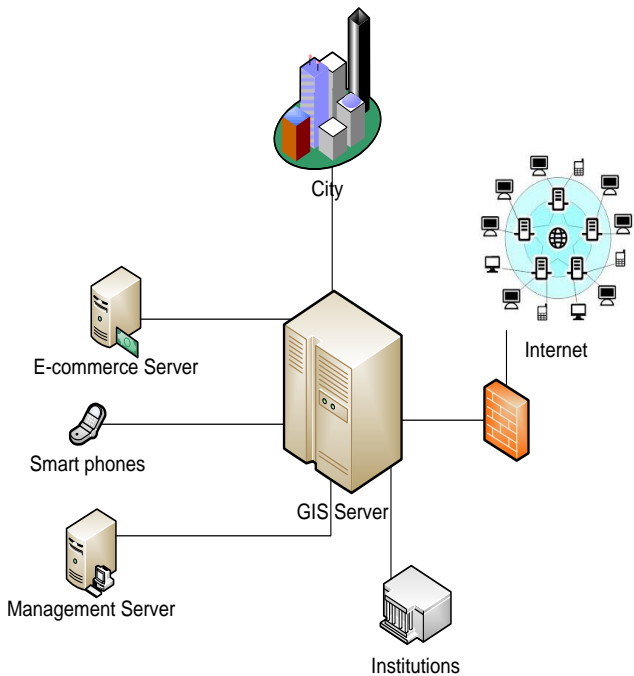

**Figure 3. Integrated GIS and MkIS System**

The GIS server connects with the various components represented as objects (City, E-commerce server, Smart

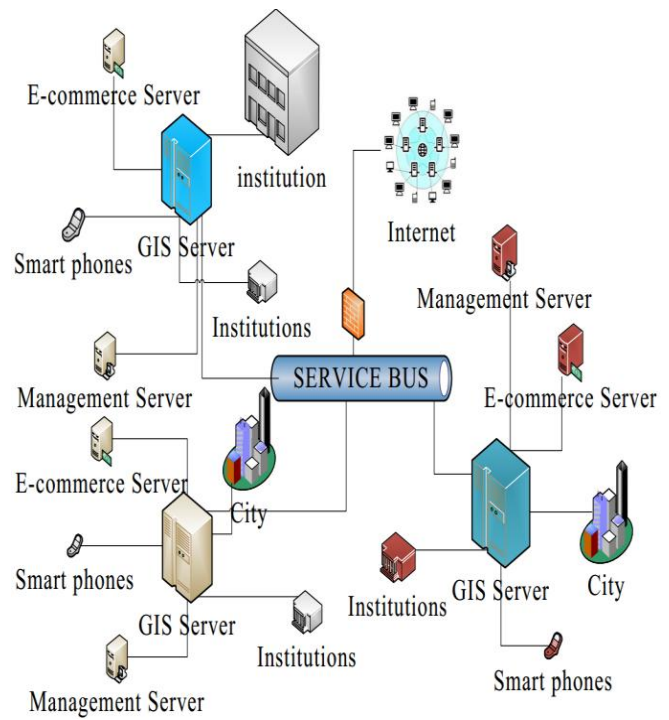

**Figure 4. Interconnecting GIS and MKIS systems via a Service Bus**

The Service Bus integrates the GIS servers across various regions. This creates a central link allowing organization and users of the various sections and organizations to be able to communicate and access inter-regional geographic, spatial and demographic data needed to promote productivity and decision making.

The centralized service bus links and shares information and data across all GIS servers nationwide. This promotes the trading of goods and services, linking retailers to the appropriate customers across the country; as GIS possesses the capacity to provide target group data as well as geographic data linking customers to their needed suppliers and vice versa.

Organizations will be able to map out the appropriate routes to use in order to provide efficient consumer delivery and increase customer satisfaction as well as productivity.

# Conclusion

The GIS technology now as a component of marketing information systems does not only provide effective way of using geodata in marketing intelligence systems but it also provides the possibilities for a better and more organized analysis of information, which is a prerequisite for making quality decisions.

With the proposition of a methodology and mechanism, integrating data mining techniques and expert's knowledge, GIS adoption will benefit organizations in many ways including improved organizational efficiency, effectiveness and better alternatives to existing practices in the market research industry. This model will now interconnect organizations at the highest levels, providing reassurance to enable broad scope of checks and balances as well as benefiting many business activities including operational, tactical and strategic decision making due to its analytical and solution driven functions.

# References

[1A. Koski, "Lossless ECG Coding," Computer Methods and Programs in Biomedicine, Vol. 52, No. 1 pp. 23–33, 1997.

[1] Marketing Research, American Marketing Association http://www.marketingpower.com/AboutAMA/Pages/DefinitionofMarketing.aspx, retrieved September 15, 2012

[2] Rethinking Marketing Research in the Digital Age http://www.smeal.psu.edu/cdt/ebrcpubs/res_papers/1999-01.html, retrieved September, 2012.

[3] World Economic Forum, "Global Information technology report", 2012, p9.

[4] Gill, H., P. Rao, "The Official Computing Guide to Data Warehousing", Que Corporation, USA, 1996.

[5] Data Warehousing, http://www.system-services.com, retrieved August 29, 2012.

[6] Inmon B. (1998). The Operational Data Store: Designing the Operational Data Store. DM Review.

[7] Hess, R.L., Rubin, R.S., and West, L.A., "Geographic information systems as a marketing information system technology". Decision Support Systems, 2004Vol. 38, Issue 2, pp. 197-212.

[8] Venkatesh.J et al, "GIS in Indian Retail Industry-A Deliberate Tool", IJCSITS Vol. 2, No.3, June 2012.

[9] ESRI news, Reliance Infocom Implements Enterprise GIS Customized to Meet Its Needs. http://www.esri.com/news/arcnews/fall02articles/reliance-infocom.html, retrieved September 12, 2012.

[10] ESRI, "GIS technology in Europe", GIS for Cadastre Management, 2005, pp2.

[11] Demographic Analysis 2009, http://www.hd.gov/HDdotGov/detail.jsp?ContentID=300, retrieved September 12, 2012.

[12] Grooms F.T, "Establishing the Foundation for Market Intelligence Worldwide usage study of the term market intelligence", Thesis,1998 pp.416

[13] Academy of Market Intelligence, http://www.mkintel.org/category/updates/, retrieved September 12, 2012.

[14] Child, J., "Information technology, organization, and the response to strategic challenges", California Management Review, Vol. 30 No. 1, pp. 33-50.

[15] Data warehouse Information Centre, http://www.dwinfocenter.org/defined.html, retrieved September 12 2012.

[16] Trubint, N., Ljubomir ,O. and Nebojša ,B, "Determining an optimal retail location by using GIS", Yugoslav Journal of Operations Research, Vol.16, No. 2, pp. 253-264

[17] Simkin, L.P. , "How retailers put location Techniques into operation", Retail and Distribution Management, pp. 21-6.

[18] Moloney, T. "Retailers Large and Small Jump on GIS Bandwagon", Computing Canada, Jan., Vol. 18, Issue 1, pp.26.

[19] QA Kester et al (2012), "Using Web Services Standards for Dealing with Complexities of Multiple Incompatible Application", IJITCS, vol.4, Issue11, pp.34-41

# Biographies

**QUIST-APHETSI KESTER** is a global award winner 2010 (First place Winner with Gold), in Canada Toronto, of the NSBE's Consulting Design Olympiad Awards and has been recognized as a Global Consulting Design Engineer. He is a PhD student in Computer Science. The PhD program is in collaboration between the AWBC/USFC Academics Without Borders/Universitaires Sans Frontieres (formerly AHED-Academics for Higher Education and Development) Canada and the Department of Computer

Science and Information Technology (DCSIT), University of Cape Coast. He had a Master of Software Engineering degree from the OUM, Malaysia and BSC in Physics from the University of Cape Coast-UCC Ghana.

He has worked in various capacities as a peer reviewer for IEEE ICAST Conference, lecturer and Head of Computer science department. He is currently a lecturer and Head of Digital Forensic Laboratory Department at the Ghana Technology University. He may be reached at kquistaphetsi@gtuc.edu.gh.

**DR. KOUMADI, KOUDJO M** is a 2009 Wilkes Award Winner BCS. He is the director of the International Institute of Technology and Management (IITM), Togo and a lecturer at Ghana Technology Graduate School. He is a member of IEEE He had his PhD in Telecommunication Engineering at the Advanced Institute of Science and Technology (KAIST), in Daejeon Korea. Masters in Telecommunication Engineering at the Advanced Institute of Science and Technology (KAIST), in Daejeon Korea and B.Sc in Telecommunication Engineering Beijing University of Posts and Telecommunications in Beijing, China. He may be reached at kkoumadi@gtuc.edu.gh.

**PROF. NII NARKU QUAYNOR is** a Jonathan B. Postel Service Award Winner 2007 from the IETF (The Internet Engineering Task Force) and a scientist and engineer who has played an important role in the introduction and development of the Internet throughout Africa. He is a Professor at the University of Cape Coast University, Ghana. He was the director of ICANN for the African Region in 2000.Quaynor graduated in engineering science from Dartmouth College in 1972 and received a Bachelor of Engineering degree from the Thayer School of Engineering there in 1973. He then studied Computer Science, obtaining an M.S. from the State University of New York at StonyBrook in 1974 and a Ph.D. from the same institution in 1977. He was founding chairman of AfriNIC, a member of the United Nations Secretary General Advisory Group on ICT, member of the ITU Telecom Board, Chair and of the OAU Internet Task Force, President of the Internet Society of Ghana, and member of the Worldbank Infodev TAP.